\documentclass[final,number,sort&compress,5p,times,twocolumn]{elsarticle}
\usepackage{lineno}

\linenumbers
\usepackage{graphicx}
\usepackage{amssymb}
\usepackage{caption}
\usepackage{subcaption}

\journal{Nuclear Instruments and Methods A}

\begin{document}

\begin{frontmatter}



\title{Efficiency and uniformity measurements of a light concentrator in combination with a SiPM array}
	
% if there is only one institution, use this form:
%\author{John Author, Giovanna Autore}
%\address{University of Wisdom, Physics City, Scienceland}

% else, use optional labels to link authors explicitly to addresses,
% as shown below:
\author[A,B]{Mariana Rihl}
\author[A,C]{Stefan Enrico Brunner}
\author[A,C]{Lukas Gruber} 
\author[A]{Johann Marton}
\author[A]{Ken Suzuki}
\address[A]{Stefan Meyer Institute for Subatomic Physics, Austrian Academy of Sciences, Boltzmanngasse 3, 1090 Vienna, Austria}
\address[B]{University of Vienna, Faculty of Physics, Boltzmanngasse 5, 1090 Vienna, Austria}
\address[C]{Vienna University of Technology, Faculty of Physics, Karlsplatz 13, 1040 Vienna, Austria}

\begin{abstract}
A position sensitive Cherenkov detector was built, consisting of 64 SiPMs with an active area of $3\times3$ mm$^2$ and a pixel size of $100\times100$ $\mu$m$^2$. The sensitive area is increased by a light concentrator which consists of 64 pyramid-shaped funnels. These funnels have an entrance area of $7\times7$ mm$^2$ and an exit area of $3\times3$ mm$^2$, guaranteeing a sufficient position resolution e.g. for the barrel DIRC detector of the PANDA experiment at FAIR. 
The efficiency and uniformity of the light concentrator in combination with the SiPM array was tested by scanning the array in two dimensions, using a pulsed light beam. Results of these tests and comparison with simulations are given here. 
\end{abstract}

\begin{keyword}
Silicon Photomultipliers
\sep
SiPM
\sep
light concentrator
\sep
Cherenkov detector
\sep
position sensitive photon detector

%% Find PACS codes here: http://www.aip.org/pacs/pacs2010/individuals/pacs2010_regular_edition/index.html


\end{keyword}

\end{frontmatter}


\section{Introduction}
\noindent Silicon Photomultipliers (SiPMs) are multi-pixel APDs operated in Geiger mode. This solid-state photon detection technology provides good single photon detection capability and high photon detection efficiency. Further features are their compact size, insensitivity to magnetic fields and cost efficiency, which make them suitable for many research fields that require photon detection, such as particle physics, nuclear physics or medical imaging. \\
A position sensitive Cherenkov detector was built, consisting of an array of $8\times8$ SiPMs (Hamamatsu S10931-­‐100P) with an active area of $3\times3$ mm$^2$ each and a pixel size of $100\times100$~$\mu$m$^2$. The signals are amplified with four 16 channel amplifiers that were built in-house and are based on Photonique amplifiers. In addition, a suitable light concentrator consisting of 64 pyramid-­‐shaped funnels was developed. With an entrance surface of $7\times7$~mm$^2$ and an exit surface of $3\times3$ mm$^2$, this light concentrator, which is made out of brass and coated with aluminium, increases the detection area of the module, while providing sufficient position resolution, e. g. for the barrel DIRC detector~\cite{DIRC4PANDA} at the PANDA experiment at the FAIR facility in Darmstadt~\cite{PANDALOI}. Increasing the detection area of the detector by this method gives several advantages. One essential advantage is that the signal-to-noise ratio improves by increasing the sensitive area using light focusing and keeping the dark count rate constant~\cite{S2Nratio}. Another benefit is that the number of read-out channels can be kept low, thus the module can be built very compactly.\\
In previous work, simulations for the collection efficiency were performed~\cite{Ahmed} as well as a scan with a laser beam to measure the collection efficiency of the module. However, the beam spot diameter was as large as 1 mm and the step size was $250\;\mu$m~\cite{Gruber}. These two parameters have been improved significantly in the new tests, providing a more detailed picture of the characteristics of the SiPMs and the light concentrator. Also, a scan with a finite incident angle was performed. The new data allows to further optimize the light guide.

\section{Test Setup}
\label{sec:testsetup}
\begin{figure}[hbt] 
\centering 
\includegraphics[width=\columnwidth,keepaspectratio]{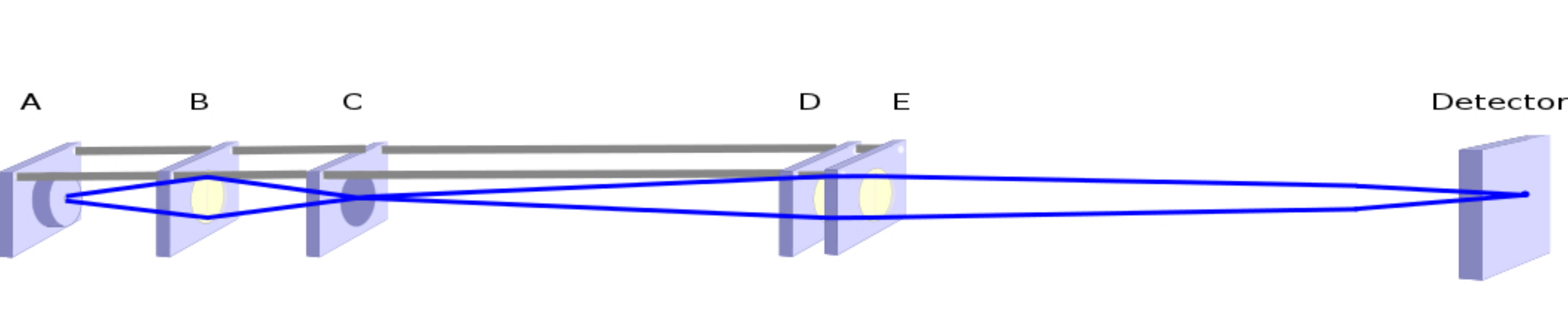}
\caption{Schematic of optomecanical items and laser beam.}
	\label{fig:schematic_laser}
		\begin{center}
		\begin{tabbing}[h]
    		\hspace*{0cm}\=\hspace*{0.85cm}\=\hspace*{0.55cm}\=\kill
	  	  ad fig.~\ref{fig:schematic_laser}.:  \> \> \textsf{A:} \> LED beam exit\\
    		\> \> \textsf{B:} \> biconvex lens with f = 30 mm \\
	   	 \> \> \textsf{C:} \> 10 $\mu$m pinhole, serves as point-like light source \\
	   	 \> \> \textsf{D:} \> collimating biconvex lens with f = 100 mm \\
		  \> \> \textsf{E:} \> focusing biconvex lens with f = 200 mm \\
    	 	 \end{tabbing}
 		\end{center}
\end{figure}

\begin{figure}[hbt] 
\centering 
\includegraphics[width=\columnwidth,keepaspectratio]{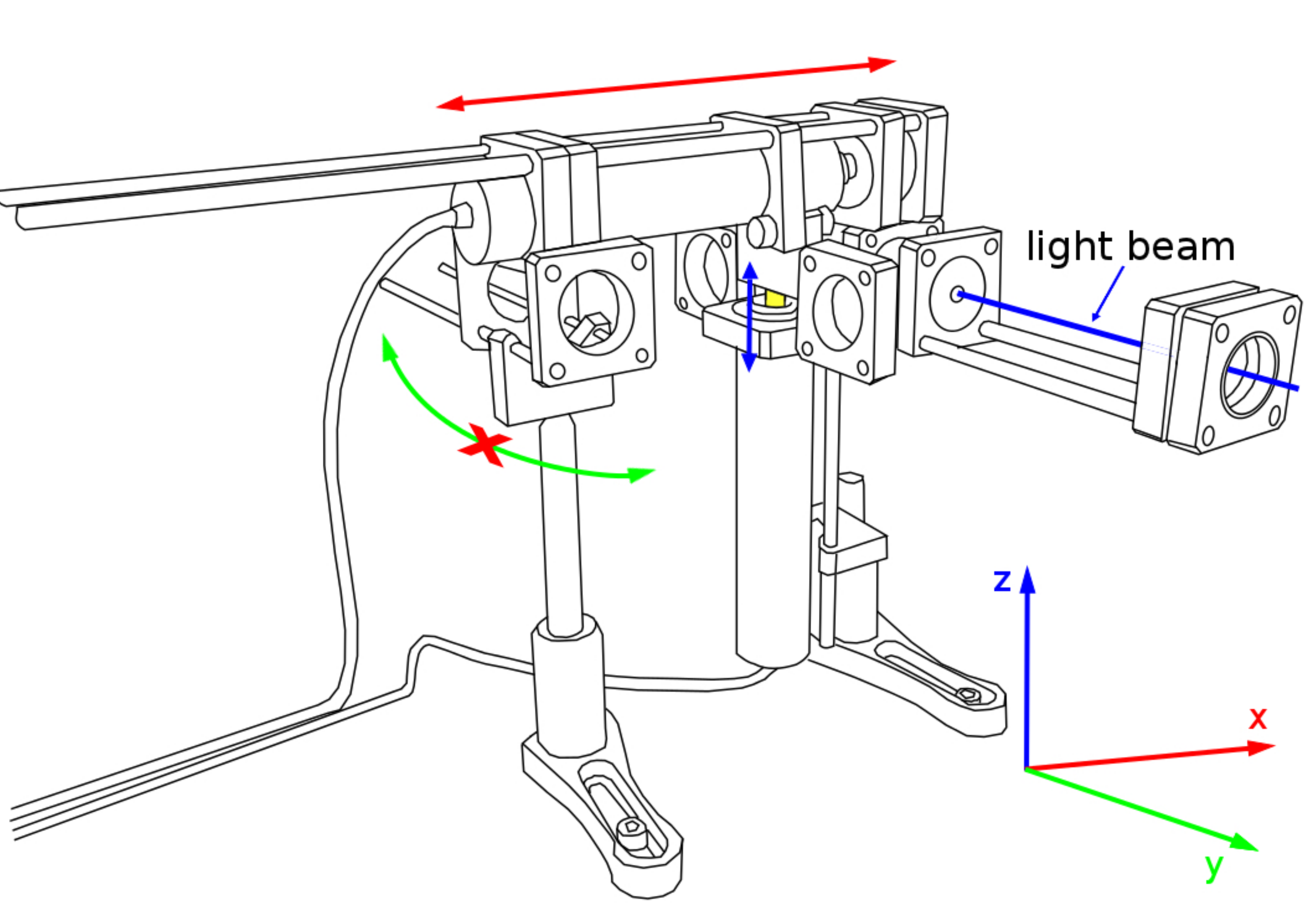}
\caption{Schematic of motor and optical setup, the coordinates x, y and z of movements are defined. }
\label{fig:motor}
\end{figure}

\noindent To test the position sensitive photon detector, the complete setup was put inside a dark box. The test setup consists of the detector module, a light source and two stepping motors which move the beam spot across the area of the scanned SiPMs. \\
The Hamamatsu 10931 $3\times3$ mm$^2$ SiPMs with a pixel size of $100\times100\;\mu$m$^2$ were chosen because they have the highest photon detection efficiency and an adequate dynamic range. The 10931 sensor series has the photon detection maximum at $\lambda=440$ nm. For the scan, a light source with a wavelength near that maximum looked reasonable and an LED with a wavelength range of $465$ nm $<\lambda<475$ nm was used. \\
The light source was set to emit pulses instead of a continuous wave in order not to saturate the sensor. 
The pulse rate of the LED was about 900 kHz with a pulse width of about 6.5 ns. \\
To reduce the beam spot diameter from $1.3\pm0.1$ mm at the LED exit to $108\pm4\;\mu$m at the SiPM surface, an optical setup, including 3 biconvex lenses and a $10\;\mu$m pinhole were included into the test setup. This optical apparatus, which is presented in figure~\ref{fig:schematic_laser} was moved by the two stepping motors, which changed the beam spot position on the detector and the array by steps of 100 $\mu$m. This guaranteed that each pixel of the SiPM was triggered by the light beam. 
\newline
\noindent During the tests, the coordinate convention was defined as follows: The x- and z-axis build a plane parallel to the detector surface and the y-axis is parallel to the beam direction. Figure~\ref{fig:motor} shows a schematic of the optical setup and its mounting on the stepping motors. Furthermore, it gives an overview of the chosen coordinate convention. 
Due to the fact that the motors are high precision tools and that the weight had to be completely poised in order to keep the precision of the motors at its high level, some measures had to be taken. The beam spot could be moved in an x- and z-direction. In order to reduce the wiggling of the motor tips, cage plates were mounted to serve as stabilisers. The optical apparatus is fixed via fixation cage plates on the x-axis motor tip, the beam direction is parallel to the y-axis of this setup. 

\begin{figure}[hbt] 
\centering 
\includegraphics[width=\columnwidth,keepaspectratio]{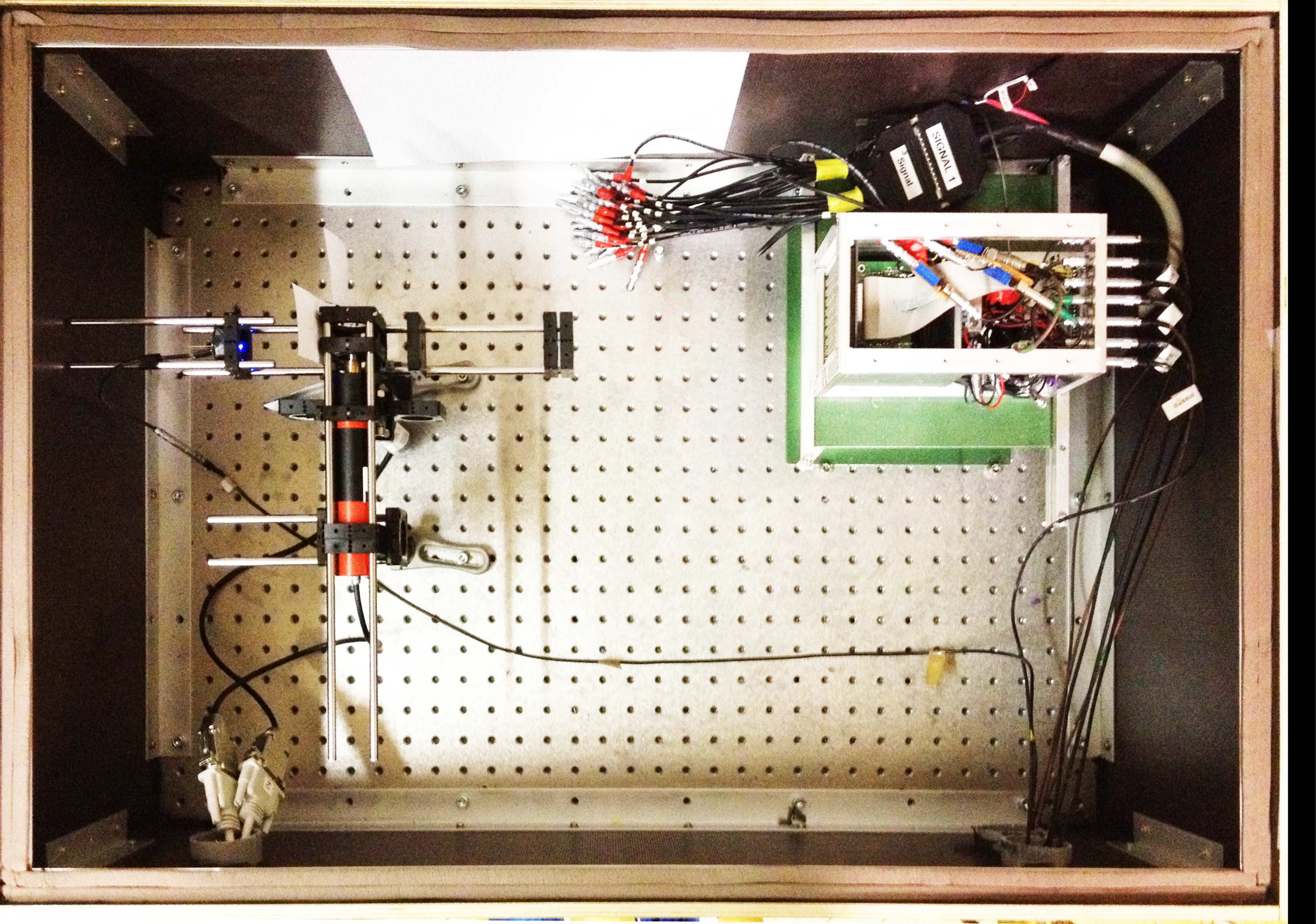}
\caption{Test setup inside dark box. On the left side of the box the optical and motor setup is mounted. On the right side of the box sits the detector prototype. }
\label{fig:darkbox}
\end{figure}

\noindent Figure~\ref{fig:darkbox} shows the opto-motoric setup together with the detector module inside the dark box. 

\section{Scanned Channels and scanning mode}
\begin{figure}[hbt] 
\centering 
\includegraphics[width=0.8\columnwidth,keepaspectratio]{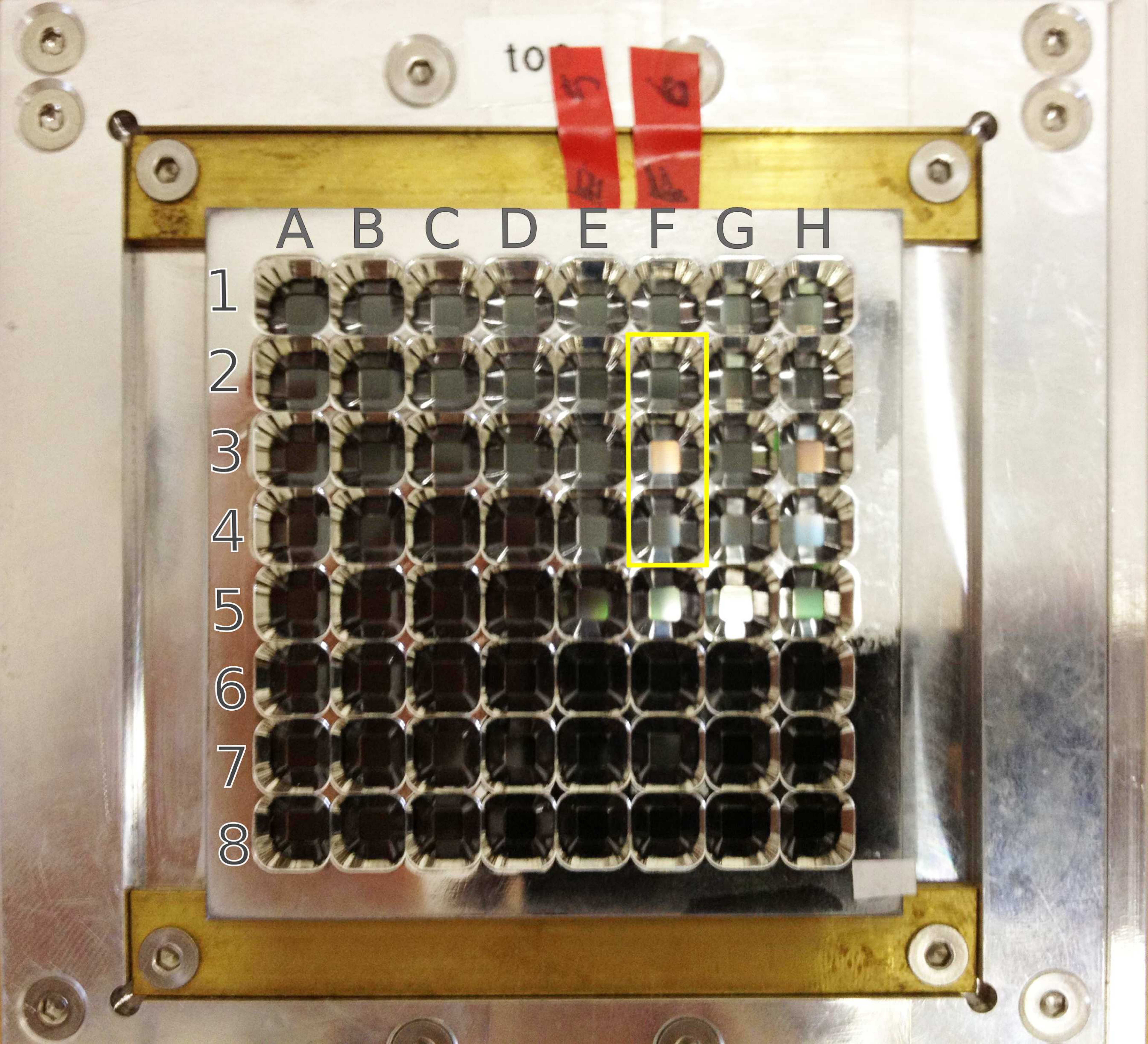}
\caption{Detector module with light concentrator. The scanned sensors are highlighted by the rectangular frame.  }
\label{fig:F2F3F4}
\end{figure}

\noindent Due to timing restraints not all 64 sensors could be scanned. Thus, three adjacent SiPMs were chosen randomly for the test. These sensors are referred to as F2, F3 and F4. Their position on the detector module surface can be seen in figure~\ref{fig:F2F3F4}. 

\noindent The sensors were scanned in three different ways. In the first two setups, all three sensors were scanned at once, with and without light concentrator. In order to test the behaviour of the collection efficiency in dependence of the incident beam angle, each sensor was scanned separately with light concentrator and an incident beam angle of about $15^\circ$.

%manual, ask your local {\it Latex} guru, or contact us at {\texttt{vci@hephy.at} .

\section{Data Acquisition}
\begin{figure}[hbt] 
\centering 
\includegraphics[width=\columnwidth,keepaspectratio]{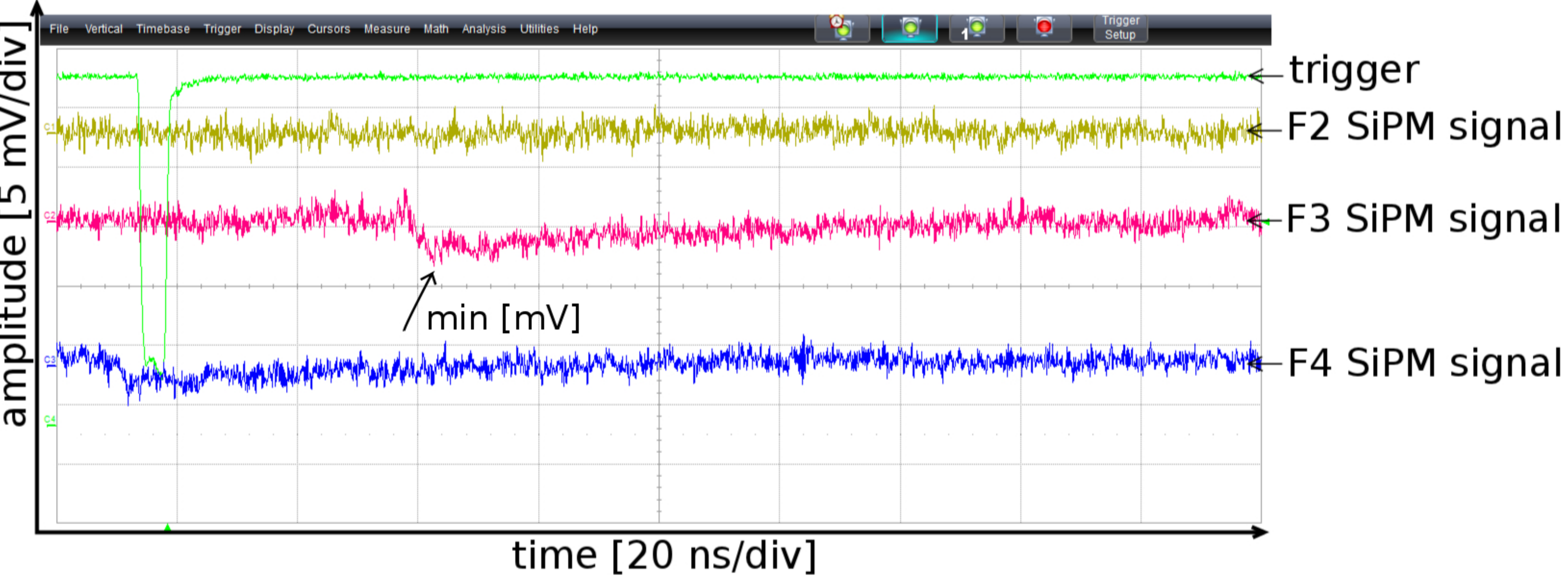}
\caption{Data acquisition with trigger and SiPM signals. Due to the beam diameter of about $108\;\mu$m (FWHM), only one SiPM sends a signal at a time, represented here by sensor F3. }
\label{fig:DAQ}
\end{figure}

%As was already mentioned in section~\ref{sec:testsetup}, the LED beam is pulsed. 
\noindent For the data acquisition, a LeCroy 735Zi WavePro digital oscilloscope was used. Three channels were used to acquire the signal, while the fourth one was used as trigger input. \\
The scope of the experiment was to extract the pulse height from the signal of the respective SiPM. The amplitude of the signal was measured by acquiring the minimum of each waveform during the acquisition window of 200 ns. To achieve good statistics, 1000 samples were taken per position of the photon source for each of the three channels respectively. The oscilloscope calculated the mean and standard deviation of 1000 samples of the amplitude. The acquired data for each channel was background corrected and then added up. The data is referred to as $\langle a\rangle_{LC}$ and $\langle a\rangle_{noLC}$ for the mean amplitude with and without light concentrator respectively. \\%dieser Absatz its 1:1 aus der Diplomarbeit übernommen und sollte zitiert bzw. abgeändert werden!
These two data values (per channel) were saved into a text file, together with information about the coordinates of the beam position. \\
Taking into account the number of data points that need to be acquired during the scans, it is obvious that an automation routine is beneficial. Such a routine was created with LabVIEW and regulates the beam spot movement by the motors as well as the data acquisition by the oscilloscope and the saving of the data. \\
Figure~\ref{fig:DAQ} shows a snapshot of the data acquisition with the oscilloscope. 
%The following example shows you how to include figures. {\it Latex} will put your figure where it considers it suitable. The publisher may subsequently still change the layout, so don't worry too much about how to position your figure. If you want to refer to a figure in the text, you should do this using its label (see Fig.~\ref{fig:VCIposter}).

\section{Results}
\subsection{Qualitative analysis}
\begin{figure}[hbt] 
\begin{center}
			\addtolength{\belowcaptionskip}{5pt}
		 \begin{subfigure}[b]{1.0\columnwidth}
           	     \centering
            	   \includegraphics[width=\columnwidth,keepaspectratio]{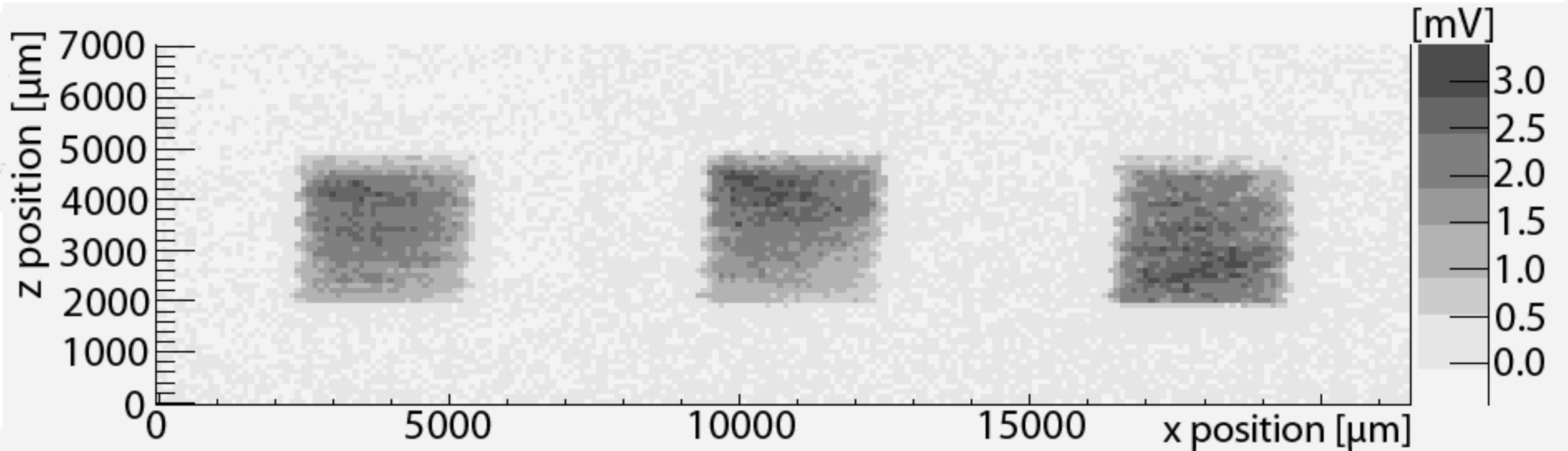}
             	   \caption{Mean intensity without light concentrator}
             	   \label{fig:TopViewa}
     		  \end{subfigure}

		 \begin{subfigure}[b]{1.0\columnwidth}
           	     \centering
            	   \includegraphics[width=\columnwidth,keepaspectratio]{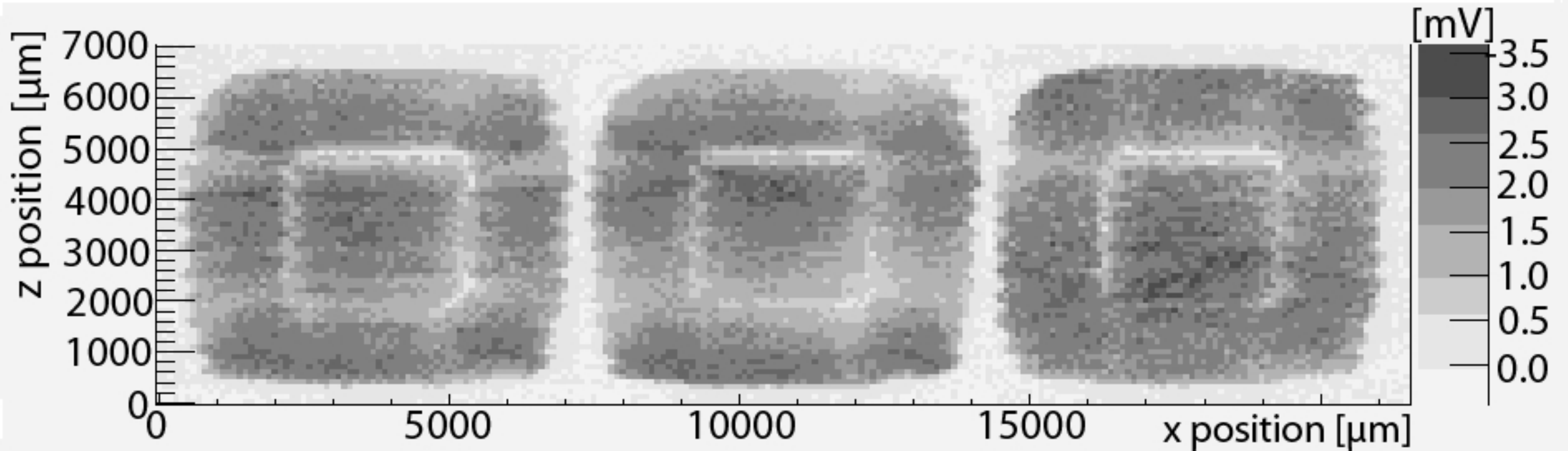}
             	   \caption{Mean intensity with light concentrator}
             	   \label{fig:TopViewb}
     		  \end{subfigure}

		 \begin{subfigure}[b]{1.0\columnwidth}
           	     \centering
            	   \includegraphics[width=\columnwidth,keepaspectratio]{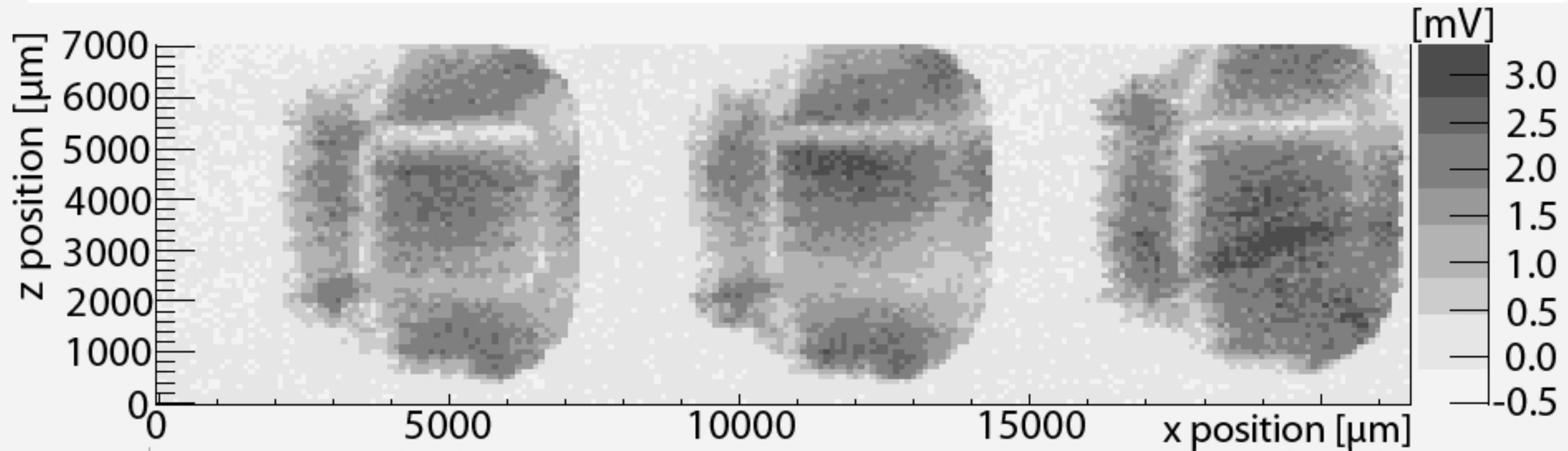}
             	   \caption{Mean intensity with light concentrator and beam angle of about $15^{\circ}$}
             	   \label{fig:TopViewc}
     		  \end{subfigure}
	\addtolength{\abovecaptionskip}{-10pt}	
	\caption{Two-dimensional histogram of the scan data for the 3 sensors (a) without LC, (b) with LC and (c) with LC and an incident beam angle of about $15^{\circ}$. The colour scheme gives the mean intensity of signal height of the SiPMs in mV. }
	\label{fig:TripleScan}
	\end{center}
\end{figure}

\begin{figure}[hbt] 
\centering 
\includegraphics[width=\columnwidth,keepaspectratio]{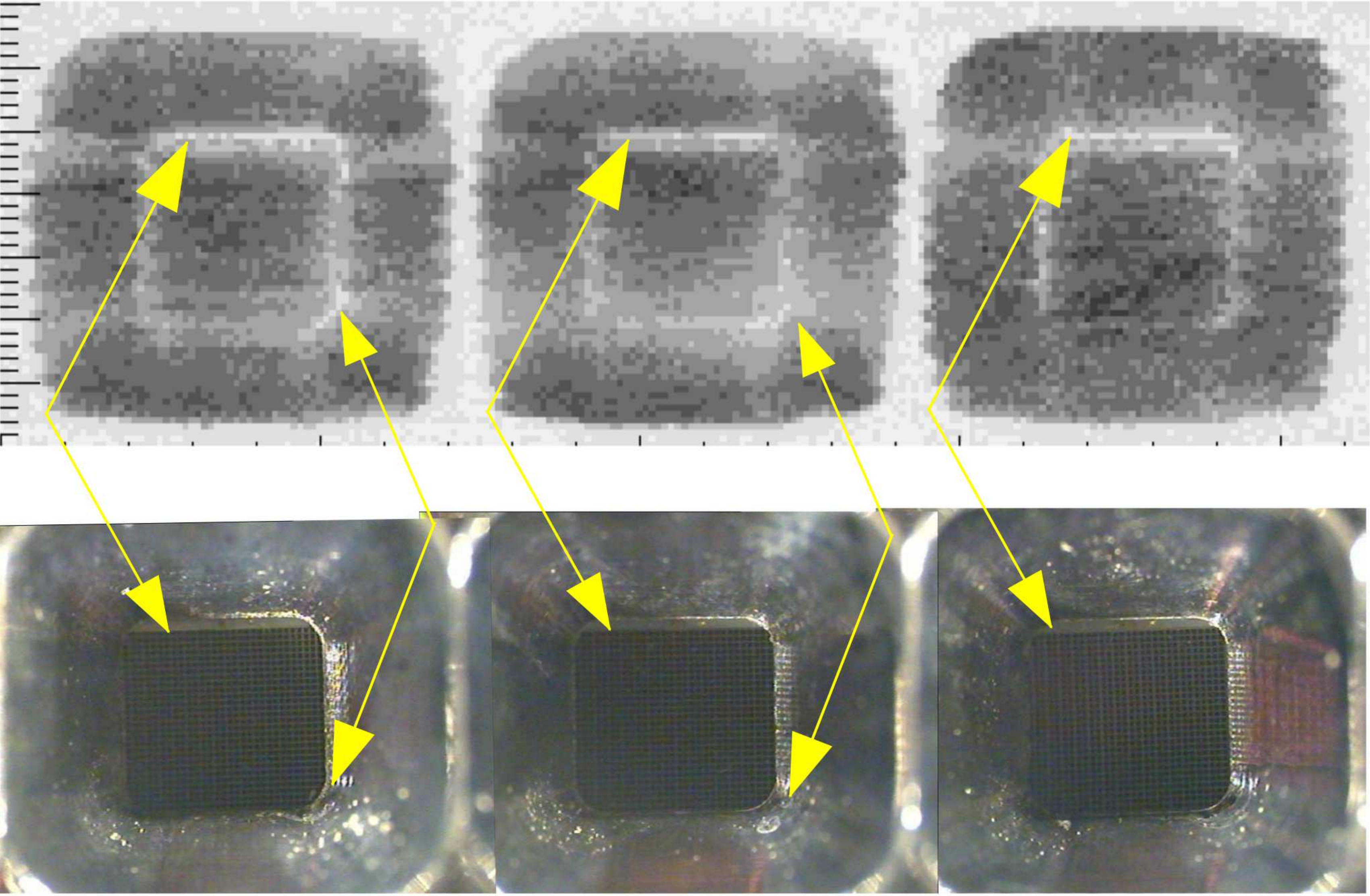}
\caption{Histogram of mean intensity and photo of the sensors with the light concentrator on top. The arrows indicate areas where no photons get detected as a result of imperfections of the alignment of the sensor array and the light concentrator. }
\label{fig:qualitative}
\end{figure}

\noindent The data, acquired during the scans was transformed into two dimensional histograms, using routines based on C++ and ROOT. Figure~\ref{fig:TripleScan} %and~\ref{fig:TripleScanSide} 
shows the two dimensional histograms from a top view. It is possible to clearly distinguish between the original sensitive area and the enhanced sensitive area when the light concentrator is applied. Also, the reduced collection efficiency due to an incident beam angle is evident in~\ref{fig:TopViewc}. 
\newline
\noindent As can be seen in figure~\ref{fig:TripleScan}, it can be distinguished between active areas and the areas where no photons get detected. One reason for the inactive area is the finite rim which separates the funnels from each other. At these areas, photons get reflected. Another reason is that the sensors were not soldered in perfect alignment, resulting in an offset between the exit area of the light concentrator and the active area of the SiPMs. Figure~\ref{fig:qualitative} shows a comparison between the two dimensional histograms and microscope photos of the respective channels, illustrating the offset of the sensors in relation to the light concentrator. 

\subsection{Collection efficiency}
\noindent The collection efficiency of the light concentrator can be calculated by comparing the data from the scans with light concentrator to the scans without the light concentrator. The collection efficiency $\epsilon_{col}$ of one funnel of the light concentrator is defined by 
\begin{equation}
\epsilon_{col}=\frac{n_d}{\alpha \cdot n_{d0}},
\label{equ:CollEffTotal}
\end{equation}
with $n_d$ being the number of photons detected with light concentrator, $n_{d0}$ the number of detected photons without light concentrator and $\alpha=(\frac{7}{3})^2\cdot0.93$ an area factor~\cite{Gruber}. The 0.93 in the area factor $\alpha$ is the geometric fill factor and puts into account the fact that the edges are rounded. \\
The area factor $\alpha$ represents the enlargement of the detection area of a SiPM and is in this specific case $A_{entrance}/A_{exit}$, where $A_{entrance}$ and $A_{exit}$ represent the entrance and exit area respectively. The collection efficiency $\epsilon_{col}$ was calculated, using the following equation for a certain funnel:\\
\begin{equation}
\epsilon_{col}=\frac{\displaystyle\sum \langle a\rangle_{LC}}{\displaystyle\sum \langle a\rangle_{{noLC}} \cdot \alpha}\\
\label{equ:CollEffSingle}
\end{equation}

%Achtung, das komplette Kapitel ist bis hier hin aus der Diplomarbeit kopiert!!!
\noindent \\Table~\ref{tab:results} shows the results for the collection efficiency for each sensor with incident beam angles of $0^\circ$ and $15^\circ$ respectively. The mean collection efficiency is also given. 
\begin{table}[htbp]
\begin{center}
\begin{tabular}{|l|c|c|}
\hline
\multicolumn{1}{|c}{Channel} &\multicolumn{1}{|c|}{Angle} &\multicolumn{1}{|c|}{Collection Efficiency {\boldmath$\epsilon_{col}$}} \\ \hline
F2  & $0^\circ$ & $88.6\;\% $   \\ 
F3 & $0^\circ$ & $83.4\;\%$  \\
F4 &  $0^\circ$ & $86.0\;\%$   \\ \hline
Mean & $0^\circ$ & $86.0\;\%$ $(\sigma=2.6\;\%)$   \\ \hline
F2 & $15^\circ$ & $56.8\;\%$   \\ 
F3 & $15^\circ$ & $55.4\;\%$ \\
F4 & $15^\circ$ & $58.4\;\%$  \\ \hline
Mean & $15^\circ$ & $56.7\;\%$ $(\sigma=1.5\;\%)$\\ \hline
\end{tabular}
\end{center}
\caption{Collection efficiencies for the evaluated three channels at two different photon incident angles. Standard deviations of the collection efficiencies are also shown, indicating the fluctuations of the collection efficiency funnel by funnel. } 
\label{tab:results}
\end{table}
\subsection{Comparison to simulations}
\begin{figure}[hbt] 
\centering 
\includegraphics[width=\columnwidth,keepaspectratio]{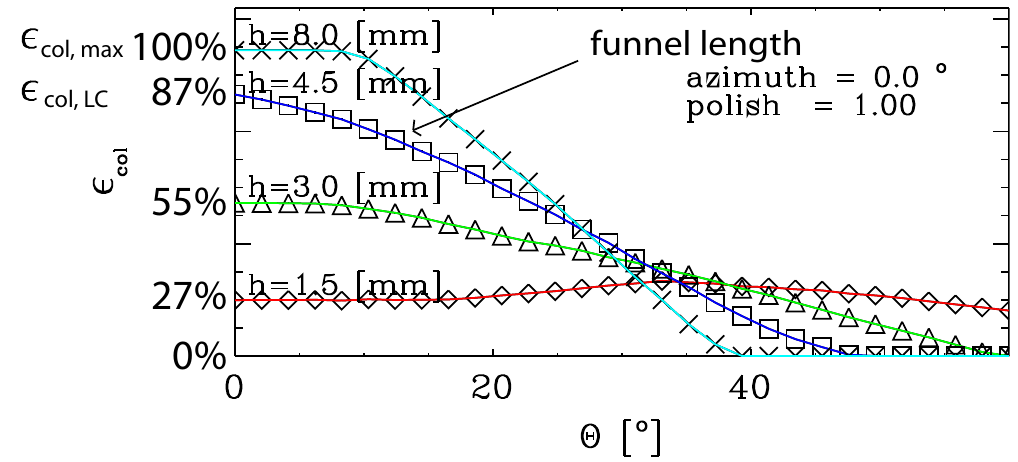}
\caption{Simulation of the collection efficiency in dependence of the incident beam angle and different funnel lengths~\cite{Ahmed}. }
\label{fig:simulation}
\end{figure}

\noindent Comparing the measured mean values with simulations of the collection efficiency of the light concentrator shows that the results are in good agreement with the simulations. The simulated collection efficiency for a light concentrator with a funnel length of $4.5$ mm and an incident beam angle perpendicular to the detector surface is about $86\;\%$. The mean of the measured collection efficiency for the light concentrator with an incident beam angle of $0^\circ$ is also about $86\;\%$. Applying an incident beam angle of $15^\circ$ results in a mean collection efficiency of about $57\;\%$, compared to the simulation value of $61\;\%$. Figure~\ref{fig:simulation} shows the results of the simulation for the light concentrator, which was done previously by the authors~\cite{Ahmed}. The figure displays the collection efficiency for different funnel lengths. The simulated collection efficiencies are given in dependence of the incident beam angle.

\section{Conclusion and outlook}
\noindent A prototype of a position sensitive SiPM array with a light concentrator was tested in order to evaluate the collection efficiency by scanning with a narrowly-focused LED light. The scans were performed with a light source of a beam spot diameter of $108\pm4\;\mu$m and a stepping size of $100\;\mu$m. These parameters have been improved significantly to earlier tests, giving a more detailed picture of the collection efficiency and uniformity. 
In addition, the performance of the light concentrator collection efficiency was tested for two different incident light beam angles, $0^\circ$ and $15^\circ$. %The data gives a good overview about the uniformity of the scanned funnels of the light concentrator. \\
%The measured collection efficiency is in good agreement with the simulations for both incident beam angles, $0^\circ$ and $15^\circ$, with a measured collection efficiency of $86\;\%$ and $57\;\%$ respectively. 
The simulation agrees well with the data and can be used to further optimise the geometry of the light concentrator.\\
 Ideas to optimise the detector include better alignment of the sensors to the concentrator or a slightly narrower exit area in order to remove the gaps in-between and to develop a different kind of light concentrator with plexiglas cones instead of a metal grid.  
%  Optimization of LC with use of simulation. Improvements: alignment, different kind of light concentrator (plexiglas), narrower area, slightly different form of the light concentrator (corners on exit different)


\begin{thebibliography}{00}
\bibitem{DIRC4PANDA} B. Seitz, et al., Nucl. Instr. and Meth. A 628 (2011) 304­. 

\bibitem{PANDALOI} W.F. Henning, J. Phys. G: Nucl. Part. Phys. 34 (2007) 551. 

\bibitem{S2Nratio} S. Korpar, et al., Nucl. Instr. and Meth. A 610 (2009) 427. 

\bibitem{Ahmed} G. Ahmed, et al., Nucl. Instr. and Meth. A 639 (2011) 107­. 

\bibitem{Gruber} L. Gruber, et al., Journal of Instrumentation 6 (2011) C11024. 

%\bibitem{HamamatsuFact} Hamamatsu, http://sales.hamamatsu.com/as-sets/pdf/parts\_S/s10362-33series\_kapd1023e05.pdf (2009). 

%\bibitem{Hamamatsu} Hamamatsu, Opto-Semiconductor Handbook (2010) 52. 

%\bibitem{GamalThesis}G. Ahmed, PhD thesis, Al-Azhar University, Cairo, 2011. 

\end{thebibliography}
\end{document}